\documentclass[aps,prl,preprint,tightenlines,superscriptaddress,showpacs,byrevtex]{revtex4}
\usepackage{graphicx}
\usepackage{epsfig}
\usepackage{dcolumn}
\usepackage{bm}
\usepackage{color}

\newcommand{\MMS}{M_{\rm rec}^2}
\newcommand{\y}{Y(4260)}

\newcommand{\yone}{Y(4360)}
\newcommand{\ytwo}{Y(4660)}
\newcommand{\lum}{{\cal L}}
\newcommand{\eff}{\varepsilon}
\newcommand{\BR}{{\cal B}}

\newcommand{\pip}{\pi^+}
\newcommand{\pim}{\pi^-}

\newcommand{\psp}{\psi(2S)}

\newcommand{\jpsi}{J/\psi}

\newcommand{\psiftf}{\psi(4415)}
\newcommand{\EE}{e^+e^-}
\newcommand{\MM}{\mu^+\mu^-}
\newcommand{\LL}{\ell^+\ell^-}
\newcommand{\pp}{\pi^+\pi^-}

\newcommand{\ppjpsi}{\pi^+\pi^- J/\psi}

\newcommand{\pppsp}{\pi^+\pi^- \psp}

\newcommand{\beq}{\begin{equation}}
\newcommand{\eeq}{\end{equation}}
\newcommand{\bitm}{\begin{itemize}}
\newcommand{\eitm}{\end{itemize}}
\def\Journal#1#2#3#4{{#1} {\bf #2}, #3 (#4)}

\def\PRL{Phys. Rev. Lett.}
\def\PRD{Phys. Rev. D}

\begin{document}

%************************************************************
       \preprint{\vbox{ \hbox{   }
                        \hbox{Belle Preprint 2007-33}
                        \hbox{KEK   Preprint 2007-27}
                         }}
\title{\quad\\[1.0cm]
Observation of Two Resonant Structures in $\EE\to \pppsp$ \\
via Initial State Radiation at Belle}

%%% Paper:    e+ e- -> pi+ pi- psi(2S) via ISR
%%% Journal:  Physical Review Letters
%%% Contacts: X. Wang (wangxiaolong@mail.ihep.ac.cn)
%%%           C. Z. Yuan (yuancz@mail.ihep.ac.cn)
%%% Non-responding authors or those who said NO are commented out.
%%% ====================================================================
%%% Click the RELOAD button on your web browser to see the updated file.
%%% ====================================================================
%%% Use \input{author} to insert this material into your latex file.
%%%%% Force institutions to appear in alphabetical order when typeset.
\affiliation{Budker Institute of Nuclear Physics, Novosibirsk}
\affiliation{Chiba University, Chiba} \affiliation{University of
Cincinnati, Cincinnati, Ohio 45221}
%%%\affiliation{Department of Physics, Fu Jen Catholic University, Taipei}
\affiliation{Justus-Liebig-Universit\"at Gie\ss{}en, Gie\ss{}en}
\affiliation{The Graduate University for Advanced Studies, Hayama}
\affiliation{Gyeongsang National University, Chinju}
\affiliation{Hanyang University, Seoul} \affiliation{University of
Hawaii, Honolulu, Hawaii 96822} \affiliation{High Energy
Accelerator Research Organization (KEK), Tsukuba}
%%%\affiliation{Hiroshima Institute of Technology, Hiroshima}
%%%\affiliation{University of Illinois at Urbana-Champaign, Urbana, Illinois 61801}
\affiliation{Institute of High Energy Physics, Chinese Academy of
Sciences, Beijing} \affiliation{Institute of High Energy Physics,
Vienna} \affiliation{Institute of High Energy Physics, Protvino}
\affiliation{Institute for Theoretical and Experimental Physics,
Moscow} \affiliation{J. Stefan Institute, Ljubljana}
\affiliation{Kanagawa University, Yokohama} \affiliation{Korea
University, Seoul}
%%%\affiliation{Kyoto University, Kyoto}
\affiliation{Kyungpook National University, Taegu}
\affiliation{Ecole Polyt\'ecnique F\'ed\'erale Lausanne, EPFL,
Lausanne} \affiliation{University of Ljubljana, Ljubljana}
\affiliation{University of Maribor, Maribor}
\affiliation{University of Melbourne, School of Physics, Victoria
3010} \affiliation{Nagoya University, Nagoya} \affiliation{Nara
Women's University, Nara} \affiliation{National Central
University, Chung-li} \affiliation{National United University,
Miao Li} \affiliation{Department of Physics, National Taiwan
University, Taipei} \affiliation{H. Niewodniczanski Institute of
Nuclear Physics, Krakow} \affiliation{Nippon Dental University,
Niigata} \affiliation{Niigata University, Niigata}
\affiliation{University of Nova Gorica, Nova Gorica}
\affiliation{Osaka City University, Osaka} \affiliation{Osaka
University, Osaka} \affiliation{Panjab University, Chandigarh}
%%%\affiliation{Peking University, Beijing}
%%%\affiliation{University of Pittsburgh, Pittsburgh, Pennsylvania 15260}
%%%\affiliation{Princeton University, Princeton, New Jersey 08544}
%%%\affiliation{RIKEN BNL Research Center, Upton, New York 11973}
\affiliation{Saga University, Saga} \affiliation{University of
Science and Technology of China, Hefei} \affiliation{Seoul
National University, Seoul}
%%%\affiliation{Shinshu University, Nagano}
\affiliation{Sungkyunkwan University, Suwon}
\affiliation{University of Sydney, Sydney, New South Wales}
\affiliation{Tata Institute of Fundamental Research, Mumbai}
\affiliation{Toho University, Funabashi} \affiliation{Tohoku
Gakuin University, Tagajo} \affiliation{Tohoku University, Sendai}
\affiliation{Department of Physics, University of Tokyo, Tokyo}
\affiliation{Tokyo Institute of Technology, Tokyo}
\affiliation{Tokyo Metropolitan University, Tokyo}
\affiliation{Tokyo University of Agriculture and Technology,
Tokyo}
%%%\affiliation{Toyama National College of Maritime Technology, Toyama}
\affiliation{Virginia Polytechnic Institute and State University,
Blacksburg, Virginia 24061} \affiliation{Yonsei University, Seoul}
  \author{X.~L.~Wang}\affiliation{Institute of High Energy Physics, Chinese Academy of Sciences, Beijing} % IHEP
  \author{C.~Z.~Yuan}\affiliation{Institute of High Energy Physics, Chinese Academy of Sciences, Beijing} % IHEP
  \author{C.~P.~Shen}\affiliation{Institute of High Energy Physics, Chinese Academy of Sciences, Beijing} % IHEP
  \author{P.~Wang}\affiliation{Institute of High Energy Physics, Chinese Academy of Sciences, Beijing} % IHEP
% \author{K.~Abe}\affiliation{High Energy Accelerator Research Organization (KEK), Tsukuba} % KEK
  \author{I.~Adachi}\affiliation{High Energy Accelerator Research Organization (KEK), Tsukuba} % KEK
  \author{H.~Aihara}\affiliation{Department of Physics, University of Tokyo, Tokyo} % Tokyo
% \author{D.~Anipko}\affiliation{Budker Institute of Nuclear Physics, Novosibirsk} % BINP
  \author{K.~Arinstein}\affiliation{Budker Institute of Nuclear Physics, Novosibirsk} % BINP
% \author{T.~Aso}\affiliation{Toyama National College of Maritime Technology, Toyama} % Toyama
% \author{V.~Aulchenko}\affiliation{Budker Institute of Nuclear Physics, Novosibirsk} % BINP
  \author{T.~Aushev}\affiliation{Ecole Polyt\'ecnique F\'ed\'erale Lausanne, EPFL, Lausanne}\affiliation{Institute for Theoretical and Experimental Physics, Moscow} % ITEP
% \author{T.~Aziz}\affiliation{Tata Institute of Fundamental Research, Mumbai} % Tata
% \author{S.~Bahinipati}\affiliation{University of Cincinnati, Cincinnati, Ohio 45221} % Cincinnati
  \author{A.~M.~Bakich}\affiliation{University of Sydney, Sydney, New South Wales} % Sydney
% \author{V.~Balagura}\affiliation{Institute for Theoretical and Experimental Physics, Moscow} % ITEP
% \author{Y.~Ban}\affiliation{Peking University, Beijing} % Peking
% \author{S.~Banerjee}\affiliation{Tata Institute of Fundamental Research, Mumbai} % Tata
  \author{E.~Barberio}\affiliation{University of Melbourne, School of Physics, Victoria 3010} % Melbourne
% \author{M.~Barbero}\affiliation{University of Hawaii, Honolulu, Hawaii 96822} % Hawaii
% \author{A.~Bay}\affiliation{Ecole Polyt\'ecnique F\'ed\'erale Lausanne, EPFL, Lausanne} % Lausanne
% ...
% Ljubljana
% \author{N.~Zwahlen}\affiliation{Ecole Polyt\'ecnique F\'ed\'erale Lausanne, EPFL, Lausanne} % Lausanne
\collaboration{The Belle Collaboration}

\date{\today}

\begin{abstract}

The cross section for $\EE\to \pppsp$ between threshold and
$\sqrt{s}=5.5$~GeV is measured using 673~fb$^{-1}$ of data on and
off the $\Upsilon(4S)$ resonance collected with the Belle detector
at KEKB. Two resonant structures are observed in the $\pppsp$
invariant mass distribution, one at $4361\pm 9\pm 9$~MeV/$c^2$
with a width of $74\pm 15\pm 10$~MeV/$c^2$, and another at
$4664\pm 11\pm 5$~MeV/$c^2$ with a width of $48\pm 15\pm
3$~MeV/$c^2$, if the mass spectrum is parameterized with the
coherent sum of two Breit-Wigner functions. These values do not
match those of any of the known charmonium states.

\end{abstract}

\pacs{14.40.Gx, 13.25.Gv, 13.66.Bc}

\maketitle

In a recently reported study of the initial state radiation
($ISR$) process, $\EE \to \gamma_{ISR} \ppjpsi$, the BaBar
Collaboration observed an accumulation of events near
4.26~GeV/$c^2$ in the invariant-mass spectrum of
$\ppjpsi$~\cite{babay4260} that they attributed to a possible new
resonance, the $\y$. This structure was also observed by the
CLEO~\cite{cleoy} and Belle Collaborations using the same
technique~\cite{belley}; in addition, there is a broad structure
near 4.05~GeV/$c^2$ in the Belle data. In a subsequent search for
the $\y$ in the $\EE \to \gamma_{ISR} \pppsp$ process, the BaBar
Collaboration observed a different structure at $m=4324\pm
24$~MeV/$c^2$ with a width of $172\pm
33$~MeV/$c^2$~\cite{babay4324} that is neither consistent with the
$\y\to \pppsp$ peak nor with $\psiftf\to \pppsp$ decay. There are
now more observed $J^{PC}=1^{--}$ states than predicted by
potential models~\cite{swanson} in the mass region between
3.8~GeV/$c^2$ and 4.5~GeV/$c^2$; it is possible that one or more
of these new states are exotic. However, it should be noted that
other interpretations that do not require resonances have been
proposed~\cite{voloshin}.

In this Letter, we report an investigation of the $\EE \to \pppsp$
process using $ISR$ events observed with the Belle
detector~\cite{Belle} at the KEKB asymmetric-energy $e^+e^-$ (3.5
on 8~GeV) collider~\cite{KEKB}. Here $\psp$ is reconstructed in
the $\ppjpsi\to \pp\LL~(\ell=e,\mu)$ final state. The integrated
luminosity used in this analysis is 673~fb$^{-1}$. About 90\% of
the data were collected at the $\Upsilon(4S)$ resonance
($\sqrt{s}=10.58$~GeV), and the rest were taken at a
center-of-mass (CM) energy 60~MeV below the $\Upsilon(4S)$ peak.

We use the PHOKHARA event generator~\cite{phokhara} to simulate
the process $\EE \to \gamma_{ISR} \pppsp$. In the generator, one
or two photons may be emitted before forming the resonance $X$,
which then decays to $\pppsp$, with $\psp\to \ppjpsi$ and
$\jpsi\to \EE$ or $\MM$. When generating $X\to \pip\pim \psp$, a
pure $S$-wave between the $\pi\pi$ system and the $\psp$, as well
as between the $\pip$ and $\pim$ is assumed. The kinematics of $X$
decays are modelled with the $\pi\pi$ invariant mass distribution
observed in our data, while $\psp\to \ppjpsi$ events are generated
according to previous measurements~\cite{besdist}.

For a candidate event, we require six good charged tracks with
zero net charge. A good charged track has transverse momentum
greater than 0.1~GeV/$c$ and impact parameters with respect to the
interaction point of $dr < 0.5$~cm in the $r$-$\phi$ plane and
$|dz|<5~(2)$~cm in the $r$-$z$ plane for pions~(leptons). For each
charged track, information from different detector subsystems is
combined to form a likelihood for each particle species ($i$),
$\mathcal{L}_i$~\cite{pid}. Tracks with
$\mathcal{R}_K=\frac{\mathcal{L}_K}{\mathcal{L}_K+\mathcal{L}_\pi}<0.4$,
are identified as pions with an efficiency of about 95\% for the
tracks of interest. Similar likelihood ratios are formed for
electron and muon identification. For electrons from $\jpsi\to
\EE$, both tracks are required to have $\mathcal{R}_e>0.1$. For
muons from $\jpsi\to \MM$, one of the tracks is required to have
$\mathcal{R}_\mu>0.95$; in addition, if one of the muon candidates
has no muon identification (ID) information, the polar angles of
the two muon candidates in the $\pip\pim \MM$ center-of-mass
system are required to satisfy $|\cos\theta_\mu|<0.75$, based on a
comparison between data and MC simulation. The lepton ID
efficiency is about 90\% for $\jpsi\to \EE$ and 87\% for $\jpsi\to
\MM$. The detection of the $ISR$ photon is not required; instead,
we require $|\MMS|<2.0~(\hbox{GeV}/c^2)^2$, where $\MMS$ is the
square of the mass recoiling against the six charged particle
system assuming that four of them are pions and the other two are
either electrons or muons. Events with $\gamma$-conversions are
removed by requiring $\mathcal{R}_e < 0.75$ for the $\pip\pim$
tracks accompanying the $\psp$.

\begin{figure}[htbp]
\centerline{\psfig{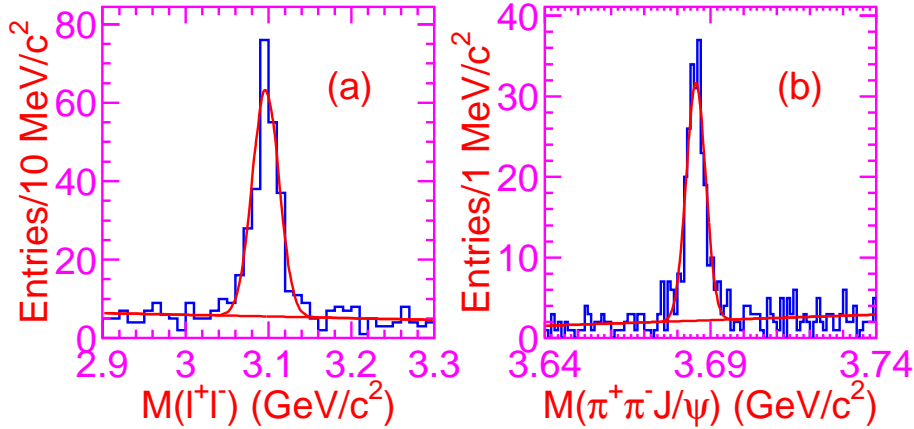}}
\caption{Invariant mass distributions of $\LL$ (a) and $\ppjpsi$
(b) for selected $\pp\pp\LL$ candidates. The curves show fits
described in the text.} \label{mllmpsp}
\end{figure}

The dilepton invariant mass distribution (the bremsstrahlung
photons in the $\EE$ final state are included) for events that
survive these selection requirements is shown in
Fig.~\ref{mllmpsp}(a); it is fitted with a Gaussian and a
second-order polynomial. A dilepton pair is considered as a
$\jpsi$ candidate if its invariant mass ($m_{\LL}$) is within $\pm
45$~MeV/$c^2$ (the mass resolution is 16~MeV/$c^2$) of the $\jpsi$
nominal mass ($m_{\jpsi}$). If there are multiple $\pp$
combinations that satisfy the $\psp$ requirements, the one with
$|m_{\pp\LL}-m_{\LL}|$, the mass difference between the $\psp$ and
$\jpsi$~\cite{PDG}, closest to $0.589~\hbox{GeV}/c^2$ is selected;
here $m_{\pp\LL}$ is the invariant mass of the $\pp\LL$ system.
Figure~\ref{mllmpsp}(b) shows the $m_{\ppjpsi}$
($=m_{\pp\LL}-m_{\LL}+m_{\jpsi}$) distribution. Fitting with a
Gaussian and a second-order polynomial yields a mass resolution of
3~MeV/$c^2$. We define a $\psp$ signal region as $m_{\ppjpsi}\in
[3.67,3.70]$~GeV/$c^2$, and a $\psp$ mass sideband region as
$m_{\ppjpsi}\in [3.64,3.67]$~GeV/$c^2$ or $m_{\ppjpsi}\in
[3.70,3.73]$~GeV/$c^2$, which is twice as wide as the signal
region.

\begin{figure}[htbp]
\psfig{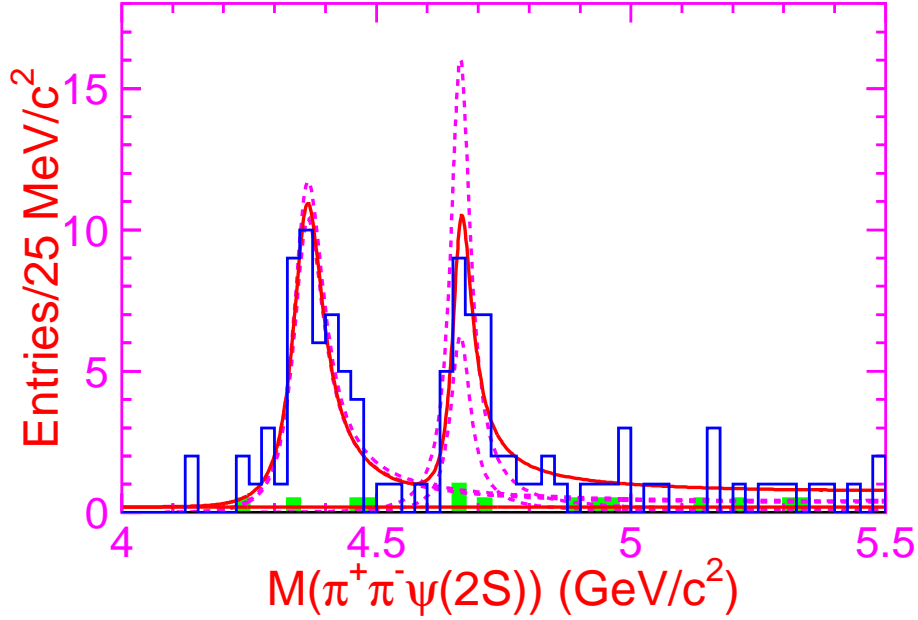}
\caption{The $\pppsp$ invariant mass distribution for events that
pass the $\psp$ selection. The open histogram is the data while
the shaded histogram is the normalized $\psp$ sidebands. The
curves show the best fit with two coherent resonances together
with a background term and the contribution from each component.
The interference between the two resonances is not shown. The two
dashed curves at each peak show the two solutions (see text).}
\label{mpppsp_full}
\end{figure}

Figure~\ref{mpppsp_full} shows the $\pppsp$ invariant mass
($m_{\pppsp}=m_{\pp\pp\LL}-m_{\pp\LL}+m_{\psp}$, where $m_{\psp}$
is the nominal $\psp$ mass) for selected $\psp$ events, together
with background estimated from the scaled $\psp$ mass sidebands.
Two distinct peaks are evident in Fig.~\ref{mpppsp_full}, one at
4.36~GeV/$c^2$ and another at 4.66~GeV/$c^2$. As can be seen from
the plot, the background determined from the $\psp$ mass sidebands
is very low. Backgrounds not described by the sidebands are
negligible; these include $\pppsp$ events, in which the $\psp$
does not decay to $\ppjpsi$ ($\jpsi\to \LL$), and events with a
$\psp$ and other particles instead of $\pp$ in the final state.

\begin{figure}[htbp]
\centerline{\psfig{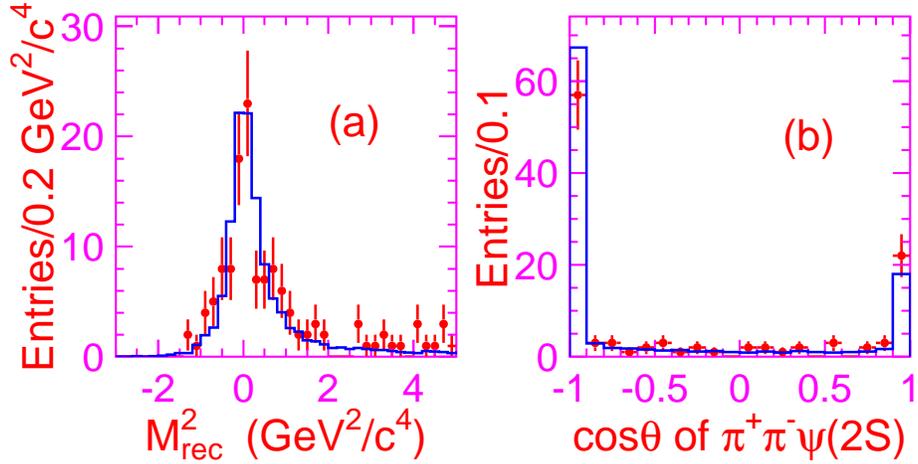}}
\caption{(a) The $\MMS$ and (b) polar angle distributions of the
$\pppsp$ system in the $\EE$ CM frame for the $\pppsp$ events with
$m_{\pppsp}\in [4.0, 5.5]$~GeV/$c^2$. The points with error bars
are data, and histograms are from MC simulation.}
\label{recmx_xcosthe_y}
\end{figure}

\begin{figure}[htbp]
\centerline{\psfig{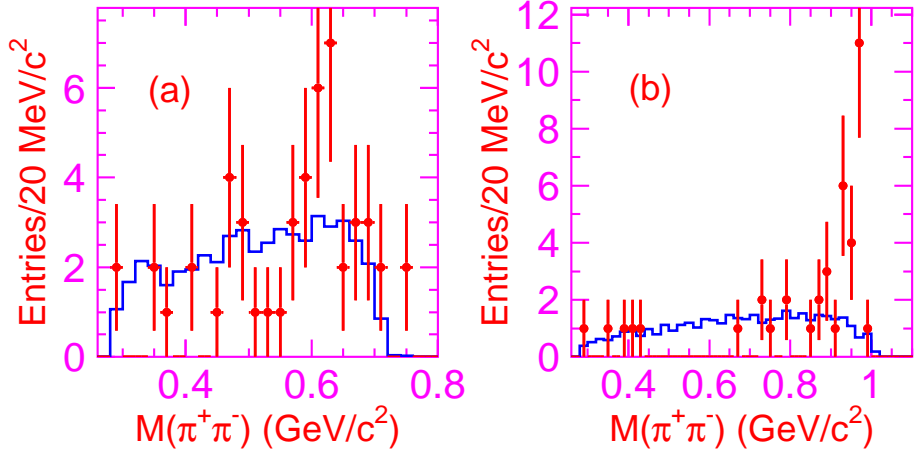}}
\caption{$\pp$ invariant mass distributions of events in different
$\pppsp$ mass regions. (a): $4.0~{\rm GeV}/c^2 <
m_{\pppsp}<4.5~{\rm GeV}/c^2$, and (b): $4.5~{\rm
GeV}/c^2<m_{\pppsp}<4.9~{\rm GeV}/c^2$. Points with error bars are
data while the histograms are MC simulation with the phase-space
distribution generated at $\sqrt{s}=4.4$~GeV (a) and 4.7~GeV (b).}
\label{mpipi_signal}
\end{figure}

Figure~\ref{recmx_xcosthe_y} shows the $\MMS$ and polar angle
distributions of the $\pppsp$ system in the $\EE$ CM frame for
$\pppsp$ events with $m_{\pppsp}\in [4.0,5.5]$~GeV/$c^2$. The data
agree with the MC simulation (shown as histograms) well,
indicating that the signal events are produced via $ISR$.
Figure~\ref{mpipi_signal} shows the $\pp$ invariant mass
distributions for events with $m_{\pppsp}\in [4.0,4.5]~{\rm
GeV}/c^2$, and $m_{\pppsp}\in [4.5,4.9]~{\rm GeV}/c^2$. In both
cases, the mass distributions differ from the phase-space
expectation and tend to be concentrated at high mass. In the high
mass resonance region, most of the $\pp$ candidates are consistent
with a $f_0(980)$ decay. %%The signal efficiency is estimated
%%by matching the $\pp$ mass distribution observed in data and is
%%9\% higher than the expectation from a phase-space model.

An unbinned maximum likelihood fit that includes two coherent
$P$-wave Breit-Wigner (BW) functions and a constant, incoherent
background is applied to the $\pppsp$ mass spectrum in
Fig.~\ref{mpppsp_full}. The BW width of each resonance is assumed
to be constant and an overall three-body phase-space factor is
applied. In the fit, the BW shapes are modified by the effective
luminosity~\cite{kuraev} and $m_{\pppsp}$-dependent efficiency,
which increases with $m_{\pppsp}$ from 3\% at 4.3~GeV/$c^2$ to 5\%
at 4.7~GeV/$c^2$. The effects of mass resolution, which is
determined from MC simulation to be 3~MeV/$c^2$-6~MeV/$c^2$ over
the full mass range, are small compared with the widths of the
observed structures, and therefore are neglected.

Figure~\ref{mpppsp_full} shows the fit results with two solutions
with equally good fit quality. In these two solutions, the masses
and widths of the resonant structures are the same, but their
partial widths to $\EE$ and the relative phase between the two
resonant structures are different (see
Table~\ref{two_sol})~\cite{beebf}. The interference is
constructive for one solution and destructive for the other. To
determine the goodness of fit, we bin the data so that the
expected number of events in a bin is at least seven and then
calculate a $\chi^2/ndf=4.7/3$ corresponding to a C.L. of 19\%.
The background level from the fit is $0.19\pm 0.14$ events per
25~MeV/$c^2$ bin, in good agreement with the $\psp$ mass sideband
estimate of $0.12\pm 0.05$. The significance of each resonance is
estimated by comparing the likelihood of fits with and without
that resonance included. We obtain a statistical significance of
more than $8\sigma$ for the first peak (hereafter referred to as
the $\yone$), and $5.8\sigma$ for the second one (the $\ytwo$).

\begin{table}
\caption{Results of the fits to the $\pp\psp$ invariant mass
spectrum. The first errors are statistical and the second
systematic. $M$, $\Gamma_{\rm tot}$, and $\BR\cdot \Gamma_{\EE}$
are the mass (in MeV/$c^2$), total width (in MeV/$c^2$), product
of the branching fraction to $\pppsp$ and the $\EE$ partial width
(in eV/$c^2$), respectively. $\phi$ is the relative phase between
the two resonances (in degrees).}\label{two_sol}
\begin{center}
\begin{tabular}{ccc}
  \hline
  Parameters & ~~~Solution I~~~ & ~~~Solution II~~~ \\
  \hline
  $M(\yone)$            & \multicolumn{2}{c}{$4361\pm 9\pm 9$} \\
  $\Gamma_{\rm tot}(\yone)$   & \multicolumn{2}{c}{$74\pm 15\pm 10$} \\
  $\BR\cdot \Gamma_{\EE}(\yone)$
                  & $10.4\pm 1.7\pm 1.5$ & $11.8\pm 1.8\pm 1.4$ \\
  $M(\ytwo)$            & \multicolumn{2}{c}{$4664\pm 11\pm 5$} \\
  $\Gamma_{\rm tot}(\ytwo)$   & \multicolumn{2}{c}{$48\pm 15\pm 3$} \\
  $\BR\cdot \Gamma_{\EE}(\ytwo)$
                  & $3.0\pm 0.9\pm 0.3$ & $7.6\pm 1.8\pm 0.8$ \\
  $\phi$          & $39\pm 30\pm 22$ & $-79\pm 17\pm 20$ \\
  \hline
\end{tabular}
\end{center}
\end{table}

The systematic errors in the mass and width measurements are
dominated by the choice of parameterization of the resonances,
especially the mass dependence of the widths; the range of changes
in the fitted values for different parameterizations is reflected
in the errors listed in Table~\ref{two_sol}. Other sources of
systematic error, such as the mass resolution and the mass scale,
are negligible.

The uncertainties in $\BR\cdot \Gamma_{\EE}$ due to the choice of
parameterization are 7\% for the $\yone$ and 10\% or 3\% for the
two $\ytwo$ solutions. There are other sources of systematic
errors for the $\BR\cdot \Gamma_{\EE}$ measurement. The particle
ID uncertainty, measured using the $\EE\to \psp\to \ppjpsi$
samples~\cite{belley}, is 5.0\%; the uncertainty in the tracking
efficiency is 1\%/track; the uncertainties in the $\jpsi$ mass,
$\psp$ mass, and $\MMS$ requirements are also measured with a
control sample of $\EE\to \psp\to \ppjpsi$ events. For these
events, the MC efficiency is found to be higher than in data by
$(4.3\pm 0.5)\%$; a correction factor is applied to the final
results and 0.5\% is taken as the associated systematic error.

Belle measures luminosity with 1.4\% precision while the
uncertainty of the radiator in the PHOKHARA program is
0.1\%~\cite{kuraev}. The main remaining uncertainty in
PHOKHARA~\cite{phokhara} is associated with the modelling of the
$\pp$ mass spectrum. A MC simulation with $\pp$ invariant mass
distributions that reflect the observations shown in
Fig.~\ref{mpipi_signal} yields an efficiency that is higher than
the phase-space simulation by about 9\%, which is used in the fits
with half of the correction (4.5\%) taken as the systematic error.
According to the MC simulation, the trigger efficiency for the
events surviving the selection criteria is around 98\% with an
uncertainty smaller than 1\%. The uncertainty in the world
average~\cite{PDG} values for $\BR(\psp\to \pp\jpsi)$ is 1.9\% and
that of $\BR(\jpsi\to \LL)=\BR(\jpsi\to \EE)+\BR(\jpsi\to \MM)$ is
1\% where we have added the errors of $\EE$ and $\MM$ modes
linearly. Finally, the statistical error in the efficiency is
1.3\%. Treating each source as independent and adding them in
quadrature, we obtain total systematic errors on $\BR\cdot
\Gamma_{\EE}$ in the range 10-14\% for the two solutions for the
$\yone$ and $\ytwo$, see Table~\ref{err_full}.

\begin{table}[htbp]
\caption{Systematic errors in the $\BR\cdot \Gamma_{\EE}$
measurement.} \label{err_full}
\begin{center}
\begin{tabular}{c c}
\hline
  Source & ~~~Relative error (\%)~~~ \\\hline
 Parameterization & 3$-$10 \\
 Particle ID &  5.0 \\
 Tracking efficiency& 6 \\
 $\jpsi$ mass, $\psp$ mass, and $\MMS$ & 0.5 \\
 Integrated luminosity & 1.4 \\
 $m_{\pp}$ distribution & 4.5 \\
 Trigger efficiency & 1 \\
 Branching fractions & 2.1 \\
 MC statistics & 1.3 \\ \hline
 Sum in quadrature & 10$-$14 \\ \hline
\end{tabular}
\end{center}
\end{table}

\begin{figure}[htbp]
\psfig{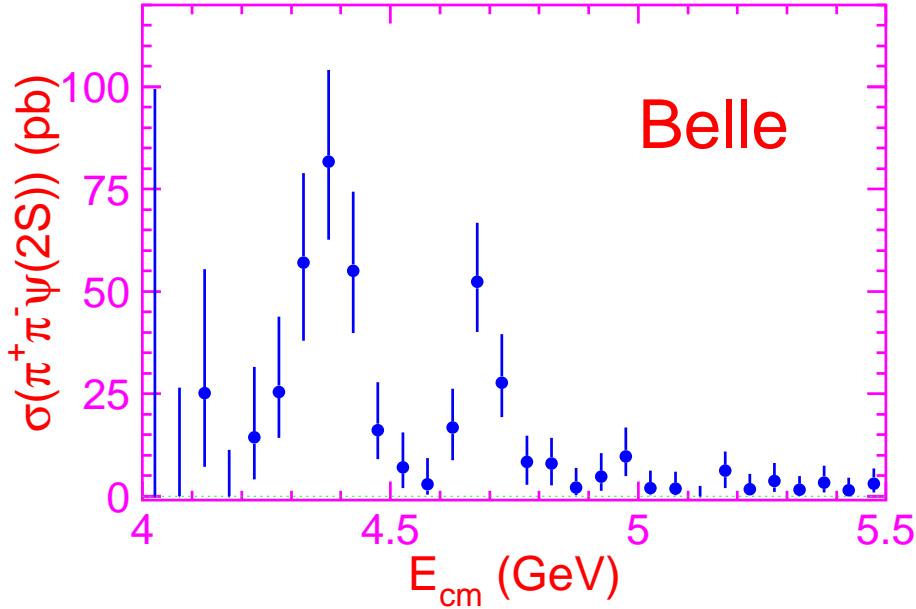} \caption{The
measured $\EE\to \pppsp$ cross section for $\sqrt{s}=4.0$~GeV to
5.5~GeV. The errors are statistical only. Bins without entries
have a central value of zero.} \label{xs_full}
\end{figure}

The cross section for $\EE\to \pppsp$ for each $\pppsp$ mass bin
is calculated according to
\[
 \sigma_i = \frac{n^{\rm obs}_i - n^{\rm bkg}}
                 {\eff_i \lum_i \BR(\psp\to \ppjpsi)\BR(\jpsi\to \LL)},
\]
where $n^{\rm obs}_i$, $\eff_i$, and $\lum_i$ are the number of
events observed in data, the efficiency, and the effective
luminosity in the $i$-th $\pppsp$ mass bin, respectively; $n^{\rm
bkg}$ is the number of background events measured in $\psp$
sidebands, taken as $0.23\pm 0.09$ events per 50~MeV/$c^2$ for all
the bins~\cite{zerobin}; $\BR(\psp\to \ppjpsi)=31.8\%$ and
$\BR(\jpsi\to \LL)=11.87\%$ are taken from Ref.~\cite{PDG}. The
resulting cross sections are shown in Fig.~\ref{xs_full}, where
the error bars include the statistical uncertainties in the signal
and the background subtraction~\cite{conrad}. The large error bars
at low mass are due to the low efficiencies. The systematic error
for the cross section measurement, which includes all the sources
listed in Table~\ref{err_full} except for that from the BW
parameterization, is 9.5\% and common to all the data points.

In summary, the $\EE\to\pppsp$ cross section is measured from
threshold up to 5.5~GeV. The measured cross sections are
consistent with results from BaBar~\cite{babay4324}. Two distinct
resonant structures are observed, one at $m=4361\pm 9\pm
9$~MeV/$c^2$ with a width of $74\pm 15\pm 10$~MeV/$c^2$,
consistent with the structure observed by BaBar in mass but with a
much narrower width, another at $m=4664\pm 11\pm 5$~MeV/$c^2$ with
a width of $48\pm 15\pm 3$~MeV/$c^2$, that has not been previously
observed. The resonant structures reported here are distinct from
the ones observed in $\EE\to\ppjpsi$~\cite{babay4260,belley}.
There are no known vector charmonium states that match these
measurements~\cite{PDG,besres}; according to potential model
calculations~\cite{eichten,godfrey}, the $4^3S_1$, $5^3S_1$, and
$3^3D_1$ charmonium states are expected to be in the mass range
close to the two resonances measured. We note that coupled-channel
effects and rescattering of pairs of charmed mesons
($D^{(*)}\bar{D}^{(*)}$, $D_s^{(*)}\bar{D}_s^{(*)}$) may affect
the above interpretation~\cite{voloshin}.

We thank the KEKB group for excellent operation of the
accelerator, the KEK cryogenics group for efficient solenoid
operations, and the KEK computer group and the NII for valuable
computing and Super-SINET network support. We acknowledge support
from MEXT and JSPS (Japan); ARC and DEST (Australia); NSFC, KIP of
CAS, and the 100 Talents program of CAS (China); DST (India);
MOEHRD, KOSEF and KRF (Korea); KBN (Poland); MES and RFAAE
(Russia); ARRS (Slovenia); SNSF (Switzerland); NSC and MOE
(Taiwan); and DOE (USA).


\begin{thebibliography}{**}
\bibitem{beebf} Considering the correlation between
$\BR(\pppsp)\cdot \Gamma_{\EE}$ and $\Gamma_{\rm tot}$, we obtain
\(\BR(\pppsp)\cdot \BR(\EE) = (14.1\pm 3.5\pm 1.4)\times 10^{-8}\)
and \((6.2\pm 2.6\pm 0.6)\times 10^{-8}\) for the $\yone$ and
$\ytwo$, respectively, for solution I; and \(\BR(\pppsp)\cdot
\BR(\EE) = (15.9\pm 3.8\pm 1.5)\times 10^{-8}\) and \((15.9\pm
4.2\pm 1.7)\times 10^{-8}\) for the $\yone$ and $\ytwo$,
respectively, for solution II.
\bibitem{zerobin} For the bins with no observed events, the number
of background events is taken to be zero when calculating the
68.3\% C.L. intervals.
\end{thebibliography}
\end{document}